# Decentralized Erasure Codes for Distributed Networked Storage

Alexandros G. Dimakis, Vinod Prabhakaran, and Kannan Ramchandran
Department of Electrical Engineering and Computer Science,
University of California, Berkeley, CA 94704.
Email: {adim, vinodmp, kannanr}@eecs.berkeley.edu

*Abstract*— We consider the problem of constructing an erasure code for storage over a network when the data sources are distributed. Specifically, we assume that there are $n$ storage nodes with limited memory and $k < n$ sources generating the data. We want a data collector, who can appear anywhere in the network, to query *any $k$ storage nodes* and be able to retrieve the data. We introduce Decentralized Erasure Codes, which are linear codes with a specific randomized structure inspired by network coding on random bipartite graphs. We show that decentralized erasure codes are optimally sparse, and lead to reduced communication, storage and computation cost over random linear coding.
Keywords: *Decentralized Erasure Codes, Wireless Networks, Network Coding, Distributed Storage.*

## I. INTRODUCTION

In this correspondence, we address the problem of distributed networked storage when there are multiple, distributed sources that generate data that must be stored efficiently in multiple storage nodes, each having limited memory. As a motivating application, one can think of sensor networks where the sensor measurements are inherently distributed and sensor motes have constrained communication, computation, and storage capabilities. In addition, distributed networked storage can be useful for peer-to-peer networks or redundant arrays of independent disks (RAID) systems. The distributed sources are $k$ data nodes, each producing one data packet of interest. We also assume we have $n$ storage nodes that will be used as a distributed network memory. If each storage node can store one data packet, we would like to diffuse the data packets so that by querying *any $k$* storage nodes, it is possible to retrieve all the $k$ data packets of interest (with high probability)[1]. The key issue, of course, is whether it is possible to achieve this robust distributed storage with minimal computation and communication.

To solve this problem, we propose *decentralized erasure codes*, which are randomized linear codes with a specific probabilistic structure that leads to optimally sparse generator matrices. These codes can be created by a randomized network protocol where each data node "pre-routes" its data packet to $O(\log n)$ randomly and independently selected storage nodes. Each storage node creates a random linear combination of whatever it happens to receive. Therefore each node operates autonomously without any central points of control and with small communication cost. In [1] the authors address the problem of distributed networked storage with a centralized server using random linear coding and demonstrate the advantages of solutions inspired by network coding for this scenario. We compare the two approaches in Section 4.

The remainder of the paper is organized as follows: In Section 2 we give the exact model assumptions and requirements and present decentralized erasure codes. In Section 3 we compare their performance with other schemes and discuss related work. In Section 4 we give some examples of using decentralized erasure codes in specific network settings and present some performance evaluation. Finally, Section 5 contains the analysis and proofs of our theorems.

## II. MODEL AND CODE CONSTRUCTION

### A. Model Assumptions

We assume that there are $k$ data-generating nodes. Without loss of generality we will assume that each data node generates one data packet. In our initial problem setting we will assume that the data packets are *independent*. For sensor network scenarios the data might be highly correlated and distributed source coding [29] can be used to compress them. We address this issue in a subsequent section. Essentially, after distributed source coding, the large correlated data packets can be replaced by smaller packets that are independent and have smaller size.

We further assume that there are $n > k$ storage nodes that will be used as storage and relay devices. We assume limited memory, and we model that by assuming that each can store only one data packet[2] (or a combination having the same number of bits as a data packet). This is a key requirement for the scalability of the network.

The ratio $k/n < 1$ is assumed fixed as $k$ and $n$ scale. For example, for a sensor network application, we can assume that some fixed ratio (for example $10\%$) of nodes are sensing while the rest are used for storage. However, as will become apparent in subsequent sections, our framework is much more general,

This research was supported by NSF under grants CCR-0219722 and CCR-0330514. This work has been presented in part in [8], [9].
The authors are with the EECS Dept., University of California, Berkeley. (Contact Mail Address: Alexandros G. Dimakis, EE Grad Student, 207 Cory Hall, U.C. Berkeley, Berkeley, CA 94720) and the $k$ data nodes and $n$ storage nodes can be any arbitrary (possibly overlapping) subsets of nodes in a larger network of $N$ nodes. We want to store the information produced in the $k$ data nodes in a redundant way in all the $n$ storage nodes. As there are $k$ data packets of interest, and each node can store no more than 1 data packet's worth of bits, it is clear that one has to query at least $k$ storage nodes to get the desired data.

[1]Throughout this correspondence, with high probability (w.h.p.) refers to at least a constant probability that can be driven arbitrarily close to one by increasing the field size $q$.

[2]More generally each storage node could store a *constant* number of packets and all the results would still hold with minor modifications

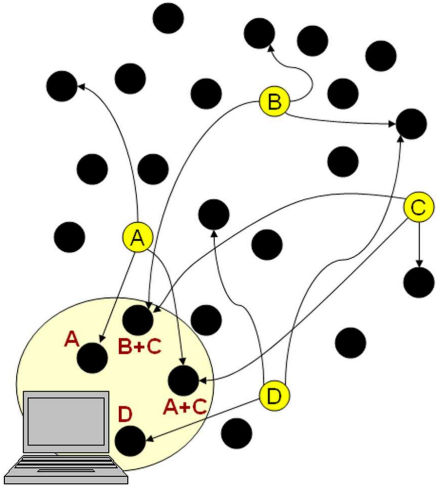

Fig. 1. Example of using linear codes for distributed storage. In this example there are $k = 4$ data nodes measuring information that is distributed and $n = 23$ storage nodes. We would like to diffuse the data to the storage nodes so that by accessing any 4 storage nodes it is possible to retrieve the data. Each data node is pre-routing to 3 randomly selected storage nodes. Each storage node only has memory to store one data packet so the ones who receive more than one packet store a linear combination of what they have received. The data collector in the example can recover the data by having access to (A, B+C, A+C, D).

The problem is to store the data in such a way that a data collector can query *any $k$ storage nodes and use the results to reconstruct the original $k$ packets* (with high probability). For instance, a data collector can get this data out of some $k$ neighboring storage nodes in its immediate vicinity to minimize the latency in a sensor network scenario. Finally we assume that the data collector has enough memory to store $k$ packets and enough processing power to run the decoding algorithm, which as we will show corresponds to solving a (sparse) system of $k$ linear equations in a finite field.

The proposed codes can be created using a randomized protocol in a network without any coordination, centralized processing or global knowledge of any sort. We assume only a network layer that can route packets from point to point (based for example on geographic information). We further assume that there are packet acknowledgements and therefore, no packets are lost. This last assumption however can be very easily relaxed due to the completely randomized nature of the solution.

Decentralized erasure codes have minimal data node degree which corresponds to a maximal sparsity of the generator matrix and minimal number of pre-routed packets. Note however, that we do not claim optimality of the distributed networked storage system as a whole. This is because we rely on a packet routing layer instead of jointly optimizing across all network layers.

*B. Decentralized Erasure Codes*

Decentralized erasure codes are random linear codes over a finite field $F_q$ with a specific randomized structure on their generator matrix. Each data packet $D_i$ is seen as a vector of elements of a finite field $f_i$. We denote the set of data nodes by $V_1$ with $|V_1| = k$ and storage nodes by $V_2$, $|V_2| = n$. We will now give a description of a randomized construction of a bipartite graph that corresponds to the creation of a decentralized erasure code. Every data node $i \in V_1$ is assigned a random set of storage nodes $N(i)$. This set is created as follows: a storage node is selected uniformly and independently from $V_2$ and added in $N(i)$ and this procedure is repeated $d(k)$ times. Therefore $N(i)$ will be smaller than $d(k)$ if the same storage node is selected twice. In fact, the size of the set $N(i)$ is exactly the number of coupons a coupon collector would have after purchasing $d(k)$ coupons from a set of $n$ coupons. It is not hard to see that when $d(k) \ll n$, $N(i)$ will be approximately equal to $d(k)$ with high probability.

Denote by $N(j) = \{i \in V_1 : j \in N(i)\}$ the set of data nodes that connect to a storage node. Each storage node will create a random linear combination of the data nodes it is connected with:

$$S_j = \sum_{\forall i : \in N(j)} f_{ij} D_i \quad (1)$$

where the coefficients $f_{ij}$ are selected uniformly and independently from a finite field $F_q$. Each storage node also stores the $f_{ij}$ coefficients, which requires an overhead storage of $N(j)(\log_2(q) + \log_2(k))$ bits.

This construction can be summarized into $s = mG$ where $s$ is a $1 \times n$ vector of stored data, $m$ is $1 \times k$ data vector and $G$ is a $k \times n$ matrix with non-zero entries corresponding to the adjacency matrix of the random bipartite graph we described. The key property that allows the decentralized construction of the code is that *each data node is choosing its neighbors independently and uniformly* or equivalently, *every row of the generator matrix is created independently* and has $N(i) = O(d(k))$ nonzero elements. This row independence, which we call "decentralized property", was proposed in our previous work [8], [9] and in [1] leads to stateless robust randomized algorithms for distributed networked storage. We compare our results with random linear coding for distributed networked storage proposed in [1] in Section 4.

A data collector querying $k$ storage nodes will gain access to $k$ encoded packets. To reconstruct, the data collector must invert a $k \times k$ submatrix $G'$ of $G$. Therefore, the key property required for successful decoding is that any selection of $G'$ forms a *full rank matrix* with high probability.

Clearly $d(k)$ is measuring the sparsity of $G$. Making $d(k)$ as small as possible is very important since it is directly related with overhead storage, decoding complexity and communication cost. Our main contribution is identifying how small $d(k)$ can be made for matrices that have the decentralized property to ensure that their submatrices are full rank with high probability. The following theorems are the main results of this correspondence:

*Theorem 1:* Let $G$ be a random matrix with independent rows constructed as described. Then, $d(k) = c\ln(k)$ is sufficient for a random $k \times k$ submatrix $G'$ of $G$ to be nonsingular with high probability. More specifically, $Pr[det(G') = 0] \leq \frac{k}{q} + o(1)$ for any $c > 5\frac{n}{k}$.

*Theorem 2:* (Converse) If each row of $G$ is generated independently (Decentralized property), at least $d(k) = \Omega(\ln(k))$ is necessary to have $G'$ invertible with high probability.

From the two theorems it follows that $d(k) = c\ln(k)$ is (order) optimal, therefore, decentralized erasure codes have minimal data node degree and logarithmically many nonzero elements in every row.

Decentralized erasure codes can be decoded using Maximum Likelihood (ML) decoding, which corresponds to solving a linear system of $k$ equations in $GF(q)$. This has a decoding complexity of $O(k^3)$. Note however that one can use the sparsity of the linear equations and have faster decoding. Using the Wiedemann algorithm [28] one can decode in $O(k^2 \log(k))$ time on average with negligible extra memory requirements.

### C. Randomized Network Algorithm

There is a very simple, robust randomized algorithm to construct a decentralized erasure code in a network. Each data node picks one out of the $n$ storage nodes randomly, pre-routes its packet and repeats $d(k) = c\ln(k)$ times. Each storage node multiplies whatever it happens to receive with coefficients selected uniformly and independently in $F_q$ and stores the result and the coefficients. A schematic representation of this is given in Figure 2.

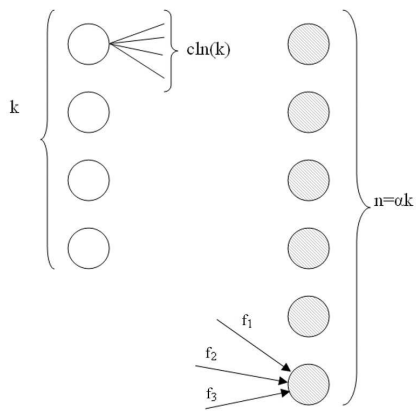

Fig. 2. Decentralized erasure codes construction. There are $d(k) = c\ln(k)$ edges starting from each data node and landing independently and uniformly on the storage nodes (If two edges have the same starting and ending point, we identify them)

### D. Storage Overhead

In addition to storing the linear combination of the received data packets, each storage node must also store the randomly selected coefficients $f_i$ [6]. The number of coefficients $N(j)$ can be determined by observing that $N(j)$ is bounded by the number of balls that land into a bin when throwing $ck\ln(k)$ balls into $n$ bins. Therefore using [26], the maximum load (the maximum number of coefficients a storage node will have to store) is $O(\log(k))$ with probability at least $1 - o(1)$. The total number of overhead bits to store the coefficients and data packet IDs is $O(\log(k)(\log(q) + \log(k)))$, which can be easily made negligible by picking larger data packet sizes. Notice that if we denote by $u = \log_2(q)$ the number of bits required to store each $f_i$, one can reduce the probability of error exponentially in the overhead bits.

## III. COMPARISON WITH OTHER SCHEMES

### A. Erasure Codes

The problem of reconstructing the $k$ packets from any $k$ out of $n$ storage nodes is essentially an erasure channel coding problem. If we assume the existence of a centralized supernode that can gather all the data, we could use any $(n, k)$ erasure code. More specifically, the centralized node would gather the $k$ data packets, use an erasure code to generate $n$ encoded packets, and assign and send one encoded packet to each storage node. If we use a good erasure code, we will be guaranteed to reconstruct the original packets by asking any $k$ encoded storage nodes. In fact, any erasure code could be used even without gathering the data in one location if there was a mechanism to create the code and coordinate the data nodes. Essentially each data node corresponds to one row in the generator matrix of the code. If that row is given to the data node (or generated using shared information), any erasure code can be created by routing each data packet to the correct storage nodes. The most common erasure codes are Reed-Solomon, which are very widely employed in many applications like computer network distributed storage systems [17], and redundant disk arrays [5]. The use of random linear codes for storage with connections to security and load balancing has been addressed in [24]. Also, LDPC codes and more recently fountain codes [19] were proposed as alternatives with randomized construction and faster encoding and decoding times. See [22] for a practical investigation on using these codes for distributed storage.

The key advantage of decentralized erasure codes is that *there is no need for coordination* among the data nodes. We show how data nodes acting randomly and independently, can create good erasure codes with sparse structure. We have completely characterized how sparse the generator matrices of these codes can be, and present a simple randomized construction algorithm that requires no centralized coordination. We believe that this property of decentralized erasure codes makes them ideal for scenarios where data is distributed and global coordination is difficult. As a side comment, note that the generator matrix of the decentralized erasure codes is never constructed explicitly and does not even exist in one place in the network.

### B. Comparison with Random Linear Coding

In [1] the authors propose the use of random linear coding inspired by network coding for distributed networked storage with one centralized server and multiple storage locations.

They compare traditional erasure codes, uncoded storage and random linear coding motivated by network coding, and demonstrate that there are significant gains in using random linear coding. In random linear coding, every element in the generator matrix of the code is selected independently and uniformly from a finite field $F_q$. This corresponds to matrices that are dense since they have a constant fraction of nonzero elements.

The main difference between our work and [1] is that we address the problem of having multiple distributed sources and no centralized server. Further, we identify how sparse can the generator matrices of decentralized codes can be, and give a simple randomized way of constructing them in a network. Sparsity leads to smaller overhead storage and more importantly, reduced communication and decoding complexity.

Specifically, random linear coding requires an overhead storage space of $O(k \log(q))$ bits, while decentralized erasure codes only $O(\log(k)(\log(q)) + \log(k)))$. The overhead storage costs are usually small if one codes over large data packets hence the communication and complexity gains are more important. If one were to use random linear coding for the multiple source networked storage problem, each data node would have to send its data to $O(n)$ storage nodes, and the total cost would be the same as flooding all the information everywhere. However using decentralized erasure codes each data node has to communicate with only $O(\log(k))$ storage nodes. As far as decoding complexity is concerned, random linear coding requires $O(k^3)$ operations to invert a dense matrix, while decentralized erasure codes can be decoded in $O(k^2 \log(k))$ by exploiting sparsity [28].

### C. Connections to Network Coding

Decentralized erasure codes can be seen as random linear network codes [14], [15], [16] on the (random) bipartite graph connecting the data and the storage nodes where each edge corresponds to one pre-routed packet. Network coding is an exciting new paradigm for communication in networks where data packets are treated as entities which can be algebraically combined rather than simply routed and stored. This fundamental idea has been initially used for maximizing multicasting throughput [2], [12] but many other advantages have been found in the recent literature [11], [21].

An equivalent way of thinking of the distributed networked storage problem is that of a random bipartite graph connecting the $k$ data nodes with the $n$ storage nodes and then adding a data collector for every possible subset of size $k$ of the $n$ storage nodes. Then the problem of multicasting the $k$ data packets to all the data collectors is equivalent to making sure that every collection of $k$ storage nodes can reconstruct the original packets. It has been shown that random linear network codes [18], [16] are sufficient for multicasting problems as long as the underlying network can support the required throughput. They key difference is that in our problem the communication graph is also random. Note that this graph does not correspond to any physical communication links but to virtual selections that are made by the randomized algorithm. Therefore this graph is not given, but can be explicitly designed to minimize communication cost. Essentially we are trying to make this random bipartite graph as sparse as possible while keeping the flow high enough and also enforcing each data node to act independently, without coordination. The theoretical analysis we give in Section 5 is based on this idea.

### D. Digital Fountain Codes

One key property of fountain codes [19], [25], is that they create every encoded packet independently and therefore have no predetermined rate (rateless property). Specifically, for LT codes, *every column* of the generator matrix is independent of the others (with logarithmic average degree, which makes it very similar to our generator matrix even though the analysis is very different). Raptor codes manage to reduce the degrees from logarithmic to constant by using an appropriate pre-code. This idea cannot be used for our problem however, since the pre-code would require centralized processing.

In this context, one can think of the decentralized property as being the transpose of the rateless property of LT codes. This is because in our case it is the rows of the generator matrix that are independent and this corresponds to *having each data source* acting independently. Our analysis is fundamentally different from the one used for fountain codes because independent columns correspond to encoded symbols being created independently. This provides fountain codes the flexibility needed to carefully design the degree distribution of the encoded symbols and make sure that belief propagation algorithms succeed. On the other hand, when the sources act independently, this enforces a degree distribution on the encoded symbols which cannot be controlled.

The decentralized property corresponds to stateless robust randomized algorithms for distributed networked storage. For sensor network applications, one implicit assumption is that it is easier for a data node to send its data to $d(k)$ randomly selected storage nodes, than it is for a storage node to find and request packets from $d'(k)$ data nodes. This is true for many practical sensor network scenarios (like the perimetric storage scenario presented in the next Section) since there will be fewer data nodes which might also be duty cycled or failing.

Another advantage is that a few sources can fail (or even be added) without affecting the performance of the code. Therefore, the proposed scheme is robust to both data and storage node failures. Schematically, each source is independently "spraying" the storage nodes with information (see also Fig. 1)) and if a collector acquires enough encoded packets, then it is possible to retrieve all the (functional) sources.

### IV. SENSOR NETWORK SCENARIOS

In this section we show how decentralized erasure codes can be applied to various sensor network scenarios and analyze their performance. It is important to realize that one can pick the $k$ data nodes and the $n$ storage nodes to be *any arbitrary subsets of nodes* of a larger network. The exact choices depend on the specific sensing application. The only requirement that we impose is that $n/k$ should remain fixed as the network scales.

In general, it is easy to determine the total communication cost involved in creating a decentralized erasure code. Each data node pre-routes to $5\frac{n}{k}\ln k$ storage nodes, therefore the total number of packets sent will be $5n\ln k$. To determine the communication cost in terms of radio transmissions, we need to impose a specific network model for routing. For example, if the diameter of the network is $D(n)$, then the total communication cost to build a decentralized erasure code will be at most $O(D(n)n\ln k)$. To become more specific we need to impose additional assumptions that depend on the specific application. If $D(n) = O(\sqrt{n})$ for example in a grid network, the total communication cost would be bounded by $O(n^{1.5}\ln k)$ to make the data available in $k = O(n)$ storage nodes.

Since each data node is essentially multicasting its packet to $O(\ln k)$ storage nodes, multicast trees can be used to minimize the communication cost. These issues depend on the specific network model and geometry and we do not address them in this paper.

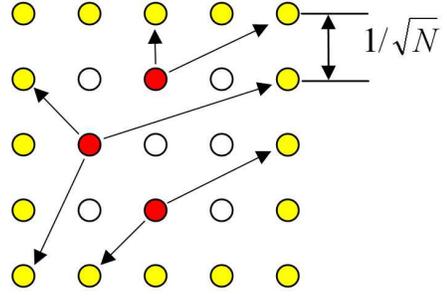

Fig. 3. Perimetric storage: The $n = 4\sqrt{N}$ nodes on the perimeter are used as storage, and $k = O(\sqrt{N})$ nodes inside the grid are the data nodes.

*A. Perimetric Storage*

To perform some experimental evaluation and also to illustrate how the decentralized erasure codes can be used as a building block for more complex applications, we consider the following scenario. Suppose we have $N$ total nodes placed on a grid in the unit square (dense scaling) and we are only interested in storing information in the $4\sqrt{N}$ nodes on the perimeter of the square (see Figure 3). This is an interesting extension since in most cases the sensor network will be monitoring an environment and potential users interested in the data will have easier access to the perimeter of this environment. Therefore we will have $n = 4\sqrt{N}$ storage nodes and $k = \rho\sqrt{N}$ data nodes for some constant $\rho < 4$. The $k$ data nodes can be placed in the grid randomly or by using some optimized sensor placement strategy [13]. Notice that we only have $O(\sqrt{N})$ nodes measuring or storing. The rest are used as relays and perhaps it is more interesting to assume that the $k$ data nodes are duty-cycled to elongate the lifetime of the network. Note that in a dense network scenario $\sqrt{N}$ can become sufficiently large to monitor the environment of interest. Again, we want to query any $k$ nodes from the perimeter and be able to reconstruct the original $k$ data packets w.h.p. The problem now is that the diameter of the network (assuming greedy geographic routing) is $O(\sqrt{N}) = O(n)$ as opposed to $\sqrt{n}$.

We assume that the transmission radius is scaling like $O(\frac{1}{\sqrt{N}})$ and measure communication cost as the total number of 1-hop radio transmissions (each transfers one packet for one hop) required to build the decentralized erasure code. It can be easily seen that the total communication cost is at most $O(N \ln N)$ which yields a logarithmic bound $O(\ln N)$ on the transmissions per node. Figure 4 illustrates some experiments on the performance under the perimetric storage scenario. Notice that the communication cost per node is indeed growing very slowly in $N$.

*B. Correlated Data*

For sensor network applications, the sensed data could be highly correlated and this correlation can be exploited to improve the performance [7], [29]. Distributed Source Coding Using Syndromes (DISCUS) [23] is a practical means of achieving this. The data nodes form the syndromes of the data packets they observe under suitable linear codes. These syndromes are treated as the data which the nodes pre-route to form the decentralized erasure codewords at the storage nodes. The data collector reconstructs the syndromes by gathering the packets from $k$ storage nodes. Using DISCUS decoding the collector can recover the original data from the syndromes. The correlation statistics, which is required by DISCUS can be learned by observing previous data at the collection point. The data nodes only need to know the rates at which they will compress their packets. This can be either communicated to them or learned adaptively in a distributed network protocol. The syndromes can be considerably shorter than the original data packets if the data observed by the different nodes are significantly correlated as is usually the case in sensor networks. Note that this approach is separating the source coding problem from the storage problem and this may not be optimal in general as shown in [27].

## V. ANALYSIS AND PROOFS

**Proof of Theorem 1**

To establish that decentralized erasure codes will be decodable, we need to show that a randomly selected square submatrix $G'$ is full rank with high probablity. For this proof we rely heavily on the Theorem 3 and use techniques similar with the ones used by Ho et al. [14], [15], [16].

It suffices to show:

$$\det G' \neq 0. \quad (2)$$

A key concept is that of a perfect matching: a bipartite graph will have a perfect matching (P.M.) if there exists a subset $E' \subseteq E$ of its edges so that no two edges in $E'$ share a common vertex and all the vertices connect to an edge in $E'$. There is a close connection between determinants of matrices and graph matchings which for the bipartite case is given by Edmonds' Theorem [20]. By construction, every row of $G'$ has a logarithmic number of non-zero coefficients chosen uniformly and independently from a finite field $F_q$. Denote these coefficients by $f_1, f_2, \cdots f_L$. Their actual number $L$ is random and approximately equal (and in fact, smaller than) $ck \ln(k)$. It suffices to show that the determinant of $G'$ is nonzero w.h.p. Note that

$$\det(G') = \sum_{\pi} sgn(\pi) \prod_{i=1}^{k} g'_{i,\pi(i)} \quad (3)$$

where we are summing over all the permutations of $\{1, 2, \cdots k\}$ and $g'_{i,j}$ is the $i, j$th element of $G'$. Notice that this is a multivariate polynomial $\det(G') = P(f_1, f_2, \cdots, f_L)$. There are two fundamentally different cases for the determinant to be zero. If for each term corresponding to each permutation there existed one or more zero elements then the determinant would be identically zero (for any choice of $f_1, f_2, \cdots f_L$). Now the key step is to notice that each permutation corresponds to exactly one potential matching of the bipartite graph. Therefore, the graph has a perfect matching if and only if $\det(G')$ is not identically zero (Edmonds' Theorem [20]). Theorem 3 establishes exactly that the random bipartite graphs we construct have perfect matchings. The other case is when $deg(G')$ is a non-zero polynomial but the specific choices of $f_1, f_2, \cdots f_L$ correspond to one of its roots. It is clear that this is a rare event and we can bound its probability using the Schwartz-Zippel Theorem [20]. Notice that the degree of $\det(G')$ is exactly $k$ when there exists a perfect matching so we obtain a bound on the probability of failure conditioned on the existence of a perfect matching: $Pr(\det(G') = 0 | \det(G') \not\equiv 0) \le \frac{k}{q}$. Which leads us to

$$Pr(\det(G') = 0) \le Pr(\det(G') \equiv 0) + \frac{k}{q}(1 - Pr(\det(G') \equiv 0)) \quad (4)$$

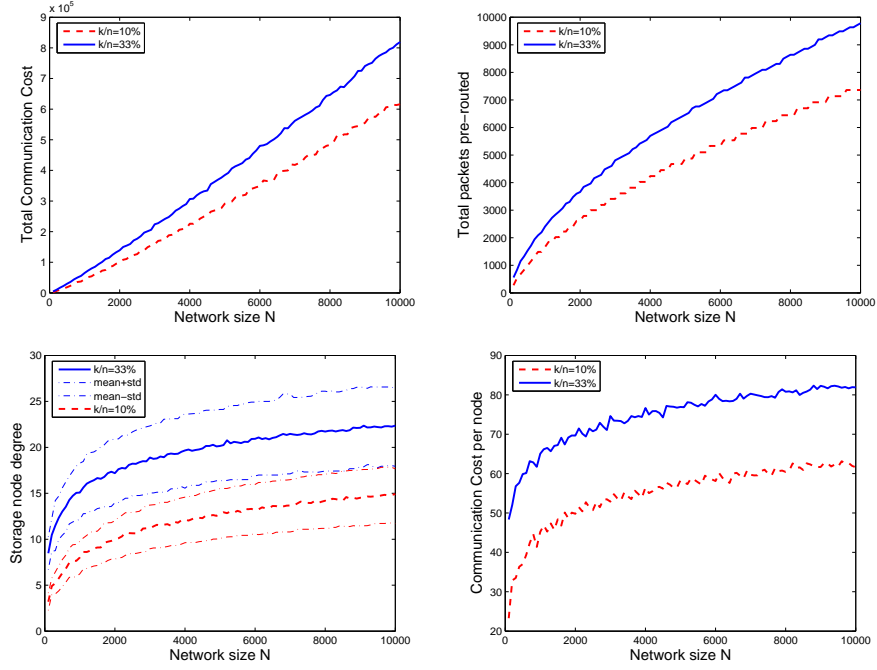

Fig. 4. Experiments for perimetric storage scenario. For each subgraph, we plot $k/n = 10\%$ (so $k = 0.4\sqrt{N}$) and $k/n = 33\%$ (so $k = 4/3\sqrt{N}$). In both cases $n = 4\sqrt{N}$
a) Total communication cost to prepare the decentralized erasure code. b) Total number of packets pre-routed. c) Average and standard deviation plots for the number of packets that are stored at storage nodes. e) Total communication cost per node.

By Theorem 3, $Pr(\det(G') \equiv 0) = o(1)$ therefore

$$Pr(\det(G') = 0) \le k/q + o(1). \quad (5)$$

∎

**Proof of Theorem 2 (Converse)** It is a standard result in balls and bins analysis [20] that in order to cover $n$ bins w.h.p. one needs to throw $\Theta(n \ln n)$ balls (See also case III in proof of Th. 3). Notice that in our case, covering all the storage nodes is necessary to have a full rank determinant (since not covering one corresponds to having a zero column in $G$). Therefore any scheme that has data nodes acting independently and uniformly will require at least $\Omega(\ln k)$ connections per data node. ∎

We have therefore demonstrated that the key technical condition we need to prove is that the random bipartite graphs we construct have a perfect matching [3] with high probability. The existence of a perfect matching guarantees that the max flow that can go through the network is sufficient. Our theoretical contribution, which may be of independent interest, is in quantifying how sparse these random bipartite graphs can be under these constraints. The proof is obtained by using an extension of a combinatorial counting technique introduced by P. Erdős and A. Rényi in [10], [4] for analyzing matchings in random bipartite graphs. The extension stems from the dependencies on the data nodes which destroy the symmetry assumed in [10], [4] thereby complicating matters.

We define the graph $B_{\ln k-left-out}$ as the random bipartite graph with two sets of vertices, $V_1, V_2$, where $|V_1| = k$, $|V_2| = n$, $n = \alpha k, (\alpha > 1)$. Every vertex in $V_1$ connects with $c \ln(k)$ vertices of $V_2$ each one chosen independently and uniformly with replacement. If two edges connect the

same two vertices we identify them. Then we pick a subset $V_2' \subset V_2$ where $|V_2'| = k$ and form the random bipartite graph $B'_{\ln k-left-out} = |V_1| \cup |V_2'|$. Edges that connect to $V_2 \setminus V_2'$ are deleted.

This graph corresponds to the submatrix $G'$ and the key property we require to establish our result is that $B'_{\ln k-left-out}$ has a perfect matching w.h.p.

*Theorem 3:* Let $B'_{\ln k-left-out}$ be a bipartite graph with $|V_1| = |V_2'| = k$ obtained from $B_{\ln k-left-out}$ by taking a random subset of $k$ storage nodes. $B'_{\ln k-left-out}$ has a perfect matching with probability $1 - o(1)$ as $k \to \infty$.

**Proof:** For a set of nodes $A \subset V_i$ of a bipartite graph B, we denote $\Gamma(A) = \{y : xy \in E(B) \text{ for some } x \in A\}$. So $\Gamma(A)$ is simply the set of nodes that connect to nodes in $A$.

A key result used in this proof is Hall's Theorem. We use it in the following form (which is easily derived from the standard Theorem [3], [4]):

*Lemma 1:* Let $B$ be a bipartite graph with vertex classes $V_1, V_2'$ and $|V_1| = |V_2'| = k$. If $B$ has no isolated vertices and no perfect matching, then there exists a set $A \subset V_i$ ($i = 1, 2$) such that:
i) $|\Gamma(A)| = |A| - 1$
ii) The subgraph $A \cup \Gamma(A)$ is connected
iii) $2 \leq |A| \leq (k+1)/2$.

The event that B has no perfect matching can be written as the union of two events. Specifically, let $E_0$ denote the event that B has one or more isolated vertices: $P(\text{B has no P.M.}) = P(E_0 \bigcup \exists A)$ (for some set $A$ satisfying Lemma (1)) Therefore by a union bound we have: $P(\text{B has no P.M.}) \leq P(E_0) + P(\exists A)$. We will treat the isolated nodes event later. We know from Lemma (1) that the size of $A$ can vary from 2 to $(k+1)/2$, so we obtain the union bound:

$$P(\exists A) = P\left(\bigcup_{i=2}^{(k+1)/2} (\exists A, |A| = i)\right) \leq \sum_{i=2}^{(k+1)/2} P(\exists A, |A| = i). \tag{6}$$

We can further partition into two cases, that the set $A$ belongs to $V_1$ (the data nodes) or $V_2'$ (the $k$ storage nodes used to decode).

$$P(\exists A) \leq \sum_{i=2}^{(k+1)/2} P(\exists A \subset V_1, |A| = i) + P(\exists A \subset V_2', |A| = i) \tag{7}$$

So we now bound the probabilities $P(\exists A \subset V_1, |A| = i)$ and $P(\exists A \subset V_2', |A| = i)$ using a combinatorial argument. *Case I: A belongs in the data nodes:* Suppose we fix $i$ nodes $A_1 \subset V_1$ and $i - 1$ nodes on $A_2 \subset V_2'$. Then the probability that a set $A = A_1$ satisfies the conditions of lemma (1) with $\Gamma(A) = A_2$ is equal to the probability that all the edges starting from $A_1$ will end in $A_2$ or are deleted. Note however that every node in $V_1$ picks $c \ln(k)$ neighbors from the set $V_2$ (which is the large set of $n = \alpha k$ nodes). We bound the probability by allowing all edges starting from $A_1$ to land in $A_2 \cup V_2 \setminus V_2'$. Therefore we have $ci \ln(k)$ edges that must land in $A_2 \cup V_2 \setminus V_2'$ and $|A_2 \cup V_2 \setminus V_2'| = i - 1 + (\alpha - 1)k$. Note that all the other edges can land anywhere and that would not affect $|\Gamma(A)|$. Therefore, since there are $\binom{k}{i}$ choices for $A_1$ and $\binom{k}{i-1}$ choices for $A_2$ we have:

$$P(\exists A \subset V_1) \leq \sum_{i=2}^{(k+1)/2} \binom{k}{i}\binom{k}{i-1}\left(\frac{i-1+(\alpha-1)k}{\alpha k}\right)^{ci \ln(k)} \tag{8}$$

We can always bound this sum by its maximum value times $k$ (since there are fewer than $k$ positive quantities added up). Therefore it suffices to show that

$$kP(\exists A \subset V_1, |A| = i) = o(1), \quad \forall i \in [2, (k+1)/2] \tag{9}$$

as $k \to \infty$.

From Stirling's approximation we obtain the bound [4] $\binom{k}{i} \leq \left(\frac{ek}{i}\right)^i$ and also it is easy to see that $\left(\frac{ek}{i-1}\right)^{i-1} \leq \left(\frac{ek}{i}\right)^i$ when $i \leq k$.

If we denote $\mathcal{X} = \left(\frac{i-1+(\alpha-1)k}{\alpha k}\right) < 1$ and use these two bounds we obtain:

$$P(\exists A \subset V_1, |A| = i) \leq \exp\left(\ln(k)(2i + ic \ln(\mathcal{X})) + 2i(1 - \ln(i))\right). \tag{10}$$

If we multiply by $k$ we get from (9) that it suffices to show

$$\exp\left(\ln(k)(2i + ic \ln(\mathcal{X}) + 1) + 2i(1 - \ln(i))\right) = o(1), \tag{11}$$

for all $i \in [2, (k+1)/2]$, as $k \to \infty$. Therefore, for this exponential to vanish it is sufficient to have the coefficient of $\ln k$ be negative:

$$2i + ic \ln(\mathcal{X}) + 1 < 0, \tag{12}$$

which gives us a bound for $c$:

$$c > \frac{-(1 + 2i)}{i \ln(\mathcal{X})}. \tag{13}$$

Notice that $\mathcal{X} < 1$ and therefore it is possible to satisfy this inequality for positive $c$. This bound should be true for every $i \in [2, (k+1)/2]$. So using

$$\frac{1 + 2i}{i} = \frac{1}{i} + 2 \leq \frac{5}{2}, \tag{14}$$

and

$$\mathcal{X} = \frac{i - 1 + (\alpha - 1)k}{\alpha k} \leq \frac{(k+1)/2 + (\alpha - 1)k}{\alpha k} \approx \frac{\alpha - 1/2}{\alpha}, \tag{15}$$

$$\frac{-1}{\ln(\mathcal{X})} \leq \frac{-1}{\ln(\frac{\alpha-1/2}{\alpha})}. \tag{16}$$

Therefore, a sufficient condition for $P(\exists A \subset V_1)$ to vanish is

$$c > \frac{-5}{2 \ln(\frac{\alpha-1/2}{\alpha})} \simeq 5\alpha. \tag{17}$$

*Case II: A belongs in the storage nodes:* With the same technique, we obtain a bound if the set $A$ is on the data nodes. This time we pick $A \subset V_2'$ with $|A| = i$ and we want $|\Gamma(A)| = i - 1$. So we require that all edges that connect to $A$ end in a specific set $A_2 \in V_1$. The extra requirement that $A \cup \Gamma(A)$ should be connected, further reduces the probability and is bounded away. To have $|\Gamma(A)| = A_2$, it must be the case that all the edges that start from $V_1 \setminus A_2$ land outside $A$. There are

$c(k - (i - 1)) \ln(k)$ such edges and each one lands outside $A$ with probability $\frac{\alpha k - i}{\alpha k}$. We therefore obtain the bound:

$$P(\exists A \subset V_2, |A| = i) \leq \binom{k}{i-1}\binom{k}{i}\left(\frac{\alpha k - i}{\alpha k}\right)^{c(k-(i-1))\ln(k)}, \quad (18)$$

which yields the condition for $c$:

$$c > \alpha \frac{k}{i} \frac{2i+1}{k-i} = 2\alpha \frac{k}{k-i} + \alpha \frac{k}{i(k-i)} \quad (19)$$

Now notice that this is a convex function of $i$ so the maximum is obtained at $i = 2$ or $i = \frac{k+1}{2}$. By substituting $i = 2$ and $i = \frac{k+1}{2}$ we find that these inequalities are always dominated by (17). So finally we require that $c > 5\alpha$.

*Case III: There exist no isolated nodes:* We will say that a data or storage node is isolated when it connects to no storage or data node respectively. Bounding the probability of this event $P(E_0)$ is easier to deal with. Notice that data nodes cannot be isolated by construction. The $\alpha k$ storage nodes receive totally $kc \ln(k)$ independent connections and we need to show that they are all covered by at least one data node w.h.p. Using a standard bound we obtain the following result ([20]):

Let $C$ denote the number of connections required to cover all $\alpha k$ data nodes. then

$$P[C > \beta \alpha k \ln(\alpha k)] \leq (\alpha k)^{-(\beta-1)}, \quad (20)$$

which shows that any $\beta > 1$ (we require $\beta > 5$) will suffice to cover all the data nodes with high probability.

Therefore from combining all the required bounds for $c$ we find that $c > 5\alpha = 5\frac{n}{k}$ is sufficient for the bipartite graph to have a perfect matching with high probability. ∎

## VI. Conclusions

We have proposed decentralized erasure codes and shown how they can be used to introduce reliable distributed storage. Our future work involves jointly optimizing summarization and code construction for multiresolution storage scenarios. Another interesting direction is investigating applications of decentralized erasure codes in peer-to-peer networks and distributed storage for computer systems. Other issues of interest involve investigating the effect of pre-routing a constant number of packets per data node as well as devising more efficient algorithms for decoding.